\documentstyle[preprint,aps]{revtex}

\begin{document}

\newcommand{\be}{\begin{equation}}
\newcommand{\ee}{\end{equation}}
\newcommand{\bea}{\begin{eqnarray}}
\newcommand{\eea}{\end{eqnarray}}
\newcommand{\ba}{\begin{array}}
\newcommand{\ea}{\end{array}}
\newcommand{\sprime}{^\prime}
\newcommand{\dprime}{^{\prime\prime}}
\newcommand{\tprime}{^{\prime\prime\prime}}

\preprint{\vbox{
\hbox{IFP-789-UNC}
\hbox{astro-ph/0010404} 
\hbox{October 2000}
}
}

\title{CMB with Quintessence: Analytic Approach and CMBFAST}
\author{James L. Crooks, James O. Dunn, Paul H. Frampton, Y. Jack Ng and Ryan M. Rohm}

\address{{\it Department of Physics and Astronomy}}

\address{{\it University of North Carolina, Chapel Hill, NC 27599-3255}}

\maketitle

\begin{abstract}
A particular kind of quintessence is considered, with
equation of motion $p_Q/\rho_Q = -1$, corresponding
to a cosmological
term with time-dependence $\Lambda(t) = 
\Lambda(t_0) (R(t_0)/R(t))^{P}$ which we examine 
initially for $0 \leq P < 3$.
Energy conservation is imposed, as is consistency with big-bang
nucleosynthesis, and the range of allowed $P$
is thereby much restricted to $0 \leq P < 0.2$.
The position of the first Doppler peak is
computed analytically and the result
combined with 
analysis of high-Z supernovae
to find how values of $\Omega_m$ and $\Omega_{\Lambda}$
depend on $P$. Some comparison is made to the CMBFAST public code.
\end{abstract} 

\newpage

Our knowledge of the universe has changed dramatically even in the
last five years. Five years ago the best guess, inspired partially by inflation,
 for the makeup of the present cosmological energy density 
was $\Omega_m = 1$ and $\Omega_{\Lambda} = 0$. However, the recent
experimental data on the cosmic background radiation and the
high - $Z$ ($Z$ = red shift) supernovae strongly suggest that both
guesses were wrong. Firstly $\Omega_m \simeq 0.3 \pm 0.1$. Second,
and more surprisingly, $\Omega_{\Lambda} \simeq 0.7 \pm 0.2$.
The value of $\Omega_{\Lambda}$ is especially unexpected for two reasons:
it is non-zero and it is $\geq 120$ orders of magnitude below its ``natural''
value.

\bigskip

The fact that the present values of $\Omega_m$ and $\Omega_{\Lambda}$
are of comparable order of magnitude is a ``cosmic coincidence''
if $\Lambda$ in the Einstein equation

\[ R_{\mu\nu} - \frac{1}{2} g_{\mu\nu} R = 8 \pi G_N T_{\mu\nu} + \Lambda g_{\mu\nu}
\]

\noindent is constant. Extrapolate the present values
of $\Omega_m$ and $\Omega_{\Lambda}$ back, say, to redshift
$Z = 100$. Suppose for simplicity that the universe is flat 
$\Omega_C = 0$ and that the present cosmic parameter values are 
$\Omega_m = 0.300...$ exactly and $\Omega_{\Lambda} = 0.700...$ exactly.
Then since $\rho_m \propto R(t)^{-3}$ (we can safely neglect radiation), 
we find that 
$\Omega_m \simeq 0.9999..$ and $\Omega_{\Lambda} \simeq 0.0000..$ at
$Z = 100$. At earlier times the ratio $\Omega_{\Lambda} /\Omega_m$
becomes infinitesimal.
There is nothing to exclude these values but it does introduce
a second ``flatness'' problem because, although we can argue for $\Omega_m + \Omega_{\Lambda}
= 1$ from inflation, the comparability of the present values of
$\Omega_m$ and $\Omega_{\Lambda}$ cries out for explanation.

\bigskip

In the present paper we shall consider a specific model of quintessence. In its context
we shall investigate the position of the first Doppler peak in the Cosmic
Microwave Background (CMB) analysis using results published by two of us with Rohm
earlier\cite{FNR}. Other works on the study of CMB include\cite{K,B,BTW,LSW}.
We shall explain some subtleties of the derivation given
in \cite{FNR} that have been raised since its publication mainly because the
formula works far better than its expected order-of-magnitude accuracy. 
Data on the CMB have been provided recently in
\cite{L,DK,M+,PSW,E,TZ,L+,l+} and especially in \cite{boomerang}.

The combination of the information about the first Doppler peak and the complementary
analysis of the deceleration parameter derived from observations of the high-red-shift
supernovae\cite{Perlmutter,Kirshner} leads to fairly precise values for the cosmic
parameters $\Omega_m$ and $\Omega_{\Lambda}$. We shall therefore also investigate the
effect of quintessence on the values of these parameters.

In \cite{FNR}, by studying the geodesics in the post-recombination period a formula was arrived
at for the position of the first Doppler peak, $l_1$.
For example, in the case of a flat universe with $\Omega_C = 0$
and $\Omega_M + \Omega_{\Lambda} = 1$ and for a conventional cosmological constant:

\bigskip

\begin{equation}
l_1 = \pi \left( \frac{R_t}{R_0} \right)
\left[\Omega_M  \left( \frac{R_0}{R_t} \right)^3 + \Omega_{\Lambda} \right]^{1/2}
\int_1^{\frac{R_0}{R_t}} \frac{dw}{\sqrt{\Omega_M w^3 + \Omega_{\Lambda}}}
\label{l1flat}
\end{equation}

\bigskip

\noindent If $\Omega_{C} < 0$ the formula becomes

\bigskip

\begin{equation}
l_1 = \frac {\pi}{\sqrt{-\Omega_C}} \left( \frac{R_t}{R_0} \right)
\left[\Omega_M  \left( \frac{R_0}{R_t} \right)^3 + \Omega_{\Lambda} + \Omega_C 
\left( \frac{R_0}{R_t} \right)^2 \right]^{1/2}
{\rm sin} \left( \sqrt{-\Omega_C} \int_1^{\frac{R_0}{R_t}} \frac{dw}{\sqrt{\Omega_M w^3 + \Omega_{\Lambda}}}
\right)
\label{l1open}
\end{equation}

\bigskip

\noindent For the third possibility of a closed universe with $\Omega_C > 0$ the formula
is:

\bigskip

\begin{equation}
l_1 = \frac {\pi}{\sqrt{\Omega_C}} \left( \frac{R_t}{R_0} \right)
\left[\Omega_M  \left( \frac{R_0}{R_t} \right)^3 + \Omega_{\Lambda} +5 \Omega_C 
\left( \frac{R_0}{R_t} \right)^2 \right]^{1/2}
{\rm sinh} \left( \sqrt{\Omega_C} \int_1^{\frac{R_0}{R_t}} \frac{dw}{\sqrt{\Omega_M w^3 + \Omega_{\Lambda}}}
\right)
\label{l1closed}
\end{equation}

\bigskip

\noindent The use of these formulas gives iso-$l_1$ lines on a $\Omega_M - \Omega_{\Lambda}$ plot
in $25 \sim 50 $\% agreement with the corresponding results found from computer code.
On the insensitivity of $l_1$ to other variables, see\cite{HW1,HW2}.
The derivation of these formulas was given in \cite{FNR}. Here we add some more details.

\bigskip

\noindent The formula for $l_1$ was derived from the relation $l_1 = \pi/\Delta\theta$ where
$\Delta\theta$ is the angle subtended by the horizon at the end of the recombination transition.
Let us consider the Legendre integral transform which has
as integrand a product of two factors, one
is the temperature autocorrelation function of the cosmic background 
radiation and the other factor is a Legendre polynomial of degree $l$. 
The issue is what is the lowest integer $l$ for which the two factors reinforce to create
the doppler peak? For small $l$ there is no reinforcement because the horizon
at recombination subtends a small angle about one degree
and the CBR fluctuations average to zero in the integral
of the Legendre transform.
At large $l$ the Legendre polynomial itself fluctuates with
almost equispaced nodes and antinodes.
The node-antinode spacing over which the Legendre polynomial varies 
from zero to a local maximum in magnitude
is, in terms of angle, on average $\pi$ divided by $l$. When this angle coincides with the angle
subtended by the last-scattering horizon,
the fluctuations of the two integrand factors are, for the first time with increasing $l$,
synchronized and reinforce (constructive interference) and
the corresponding partial wave coefficient is larger than for slightly smaller
or slightly larger $l$.
This explains the occurrence of $\pi$ in the equation for the $l_1$ value of the first doppler peak
written as $l_1 = \pi/\Delta\theta$.

\bigskip

Another detail concerns whether to use the photon or 
acoustic horizon, where the former is $\sqrt{3}$
larger than the latter?
If we examine the evolution
of the recombination transition given in \cite{Peebles}
the degree of ionization is 99\% at $5,000^0$K
(redshift $Z=1,850$) falling to 1\% at $3,000^0$K ($Z = 1,100$).
One can see qualitatively that
during the recombination transition the fluctuation
can grow.
The agreement
of the formula for $l_1$, using the photon horizon,
with experiment shows phenomenologcally
that the fluctuation does grow
during the recombination transition and that is why there is no full factor of 
$\sqrt{3}$, as would arise using the acoustic horizon, in its
numerator. 
When we look at the CMBFAST code below, 
we shall find a factor
in $l_1$ of $\sim1.22$, intermediate between $1$ (optical) 
and $\sqrt{3}$ (acoustic). 

\bigskip
\bigskip

To introduce our quintessence model as a time-dependent cosmological term,
we start from the Einstein equation:

\begin{equation}
R_{\mu\nu} - \frac{1}{2} R g_{\mu\nu} = \Lambda(t) g_{\mu\nu} + 8 \pi G T_{\mu\nu} = 8 \pi G {\cal T}_{\mu\nu}
\label{einstein}
\end{equation}

\noindent where $\Lambda(t)$ depends on time as will be specified later
and $T_{\nu}^{\mu} = {\rm diag} (\rho, -p, -p, -p)$.
Using the Robertson-Walker metric, the `00' component of Eq.(\ref{einstein})
is

\begin{equation}
\left( \frac{\dot{R}}{R} \right)^2 + \frac{k}{R^2} = \frac{8 \pi G \rho}{3} +
 \frac{1}{3}\Lambda
\label{00}
\end{equation}

\noindent while the `ii' component is
\begin{equation}
2\frac{\ddot{R}}{R} + \frac{\dot{R}^2}{R^2} + \frac{k}{R^2} = -8 \pi G p + \Lambda
\label{ii}
\end{equation}
Energy-momentum conservation follows from Eqs.(\ref{00},\ref{ii}) because of the Bianchi identity
$D^{\mu} (R_{\mu\nu} - \frac{1}{2} g_{\mu\nu}) = D^{\mu} (\Lambda g_{\mu\nu} + 8\pi G T_{\mu\nu})
= D^{\mu} {\cal T}_{\mu\nu} = 0$.

\bigskip

Note that the separation of ${\cal T}_{\mu\nu}$ into two terms, one involving $\Lambda(t)$,
as in Eq(\ref{einstein}), is not meaningful except in a phenomenological sense because of energy conservation. 

\bigskip

In the present cosmic era, denoted by the subscript `0', Eqs.(\ref{00},\ref{ii}) become respectively:

\begin{equation}
\frac{8\pi G}{3} \rho_0 = H_0^2 + \frac{k}{R_0^2} - \frac{1}{3} \Lambda_0
\label{00now}
\end{equation}
\begin{equation}
- 8 \pi G p_0 = - 2 q_0 H_0^2 + H_0^2 + \frac{k}{R_0^2} - \Lambda_0
\label{iinow}
\end{equation}
where we have used $q_0 = - \frac{\ddot{R}_0}{R_0 H_0^2}$ and $H_0 = \frac{\dot{R}_0}{R_0}$.

For the present era, $p_0 \ll \rho_0$ for cold matter and then Eq.(\ref{iinow}) becomes:

\begin{equation}
q_0 = \frac{1}{2} \Omega_{M} - \Omega_{\Lambda}
\label{decel}
\end{equation}
where $\Omega_{M} = \frac{8 \pi G \rho_0}{3 H_0^2}$ and $\Omega_{\Lambda} = \frac{\Lambda_0}{3 H_0^2}$.

\bigskip
\bigskip

Now we can introduce the form of $\Lambda(t)$ we shall assume by writing

\begin{equation}
\Lambda(t) = b R(t)^{-P}  
\end{equation}
where $b$ {\bf is} a constant and the exponent $P$ we shall study for the
range $0 \leq P < 3$. This motivates the introduction of the new variables

\begin{equation}
\tilde{\Omega}_M = \Omega_M - \frac{P}{3 - P} \Omega_{\Lambda} , ~~~~\tilde{\Omega}_{\Lambda}
= \frac{3}{3 - P} \Omega_{\Lambda}
\label{tilde}
\end{equation}

\noindent It is unnecessary to redefine $\Omega_C$ 
because $\tilde{\Omega}_M + \tilde{\Omega}_{\Lambda}
= \Omega_M + \Omega_{\Lambda}$. The case $P=2$ was proposed, at least for late
cosmological epochs, in \cite{chen}.

\bigskip
\bigskip
\bigskip

The equations for the first Doppler peak incorporating the possibility of non-zero $P$
are found to be the following modifications of Eqs.(\ref{l1flat},\ref{l1open},\ref{l1closed}).
For $\Omega_C=0$

\bigskip

\begin{equation}
l_1 = \pi \left( \frac{R_t}{R_0} \right)
\left[\tilde{\Omega}_M  \left( \frac{R_0}{R_t} \right)^3 + \tilde{\Omega}_{\Lambda}
\left( \frac{R_0}{R_t} \right)^P
 \right]^{1/2}
\int_1^{\frac{R_0}{R_t}} \frac{dw}{\sqrt{\tilde{\Omega}_M w^3 + \tilde{\Omega}_{\Lambda} w^P}}
\label{l1Pflat}
\end{equation}

\bigskip

\noindent If $\Omega_{C} < 0$ the formula becomes

\bigskip

\begin{eqnarray}
l_1 & = & \frac {\pi}{\sqrt{-\Omega_C}} \left( \frac{R_t}{R_0} \right)
\left[\tilde{\Omega}_M  \left( \frac{R_0}{R_t} \right)^3 + \tilde{\Omega}_{\Lambda}
\left( \frac{R_0}{R_t} \right)^P
 + \Omega_C 
\left( \frac{R_0}{R_t} \right)^2 \right]^{1/2} \times \nonumber \\
& & ~~~~ \times {\rm sin} \left( \sqrt{-\Omega_C} \int_1^{\frac{R_0}{R_t}} \frac{dw}{\sqrt{\tilde{\Omega}_M w^3 
+ \tilde{\Omega}_{\Lambda} w^P + \Omega_C w^2}}
\right)
\label{l1Popen}
\end{eqnarray}

\bigskip

\noindent For the third possibility of a closed universe with $\Omega_C > 0$ the formula
is:

\bigskip

\begin{eqnarray}
l_1 & = & \frac {\pi}{\sqrt{\Omega_C}} \left( \frac{R_t}{R_0} \right)
\left[\tilde{\Omega}_M  \left( \frac{R_0}{R_t} \right)^3 + \tilde{\Omega}_{\Lambda}
\left( \frac{R_0}{R_t} \right)^P
 + \Omega_C 
\left( \frac{R_0}{R_t} \right)^2 \right]^{1/2} \times \nonumber \\
& & ~~~ \times ~~~ {\rm sinh} \left( \sqrt{\Omega_C} \int_1^{\frac{R_0}{R_t}} \frac{dw}{\sqrt{\tilde{\Omega}_M w^3 
+ \tilde{\Omega}_{\Lambda} w^P + \Omega_C w^2}}
\right)
\label{l1Pclosed}
\end{eqnarray}

\bigskip
\bigskip

\noindent The dependence of $l_1$ on $P$ is illustrated for constant $\Omega_M = 0.3$ in
Fig. 1(a), and for the flat case $\Omega_C = 0$ in Fig. 1(b). For illustration we have varied
$0 \leq P < 3$ but as will become clear later in the paper (see Fig 3 below) only the
much more restricted range $0 \leq P < 0.2$ is possible for a fully consistent
cosmology when one considers evolution since the
nucleosynthesis era.

\bigskip
\bigskip
\bigskip
\bigskip

\noindent We have introduced $P$ as a parameter which is real and with 
$0 \leq P < 3$.
For $P \rightarrow 0$ we regain the standard cosmological model. But now
we must investigate other restrictions already necessary for $P$ before precision
cosmological measurements restrict its range even further.

\bigskip
\bigskip

Only for certain $P$ is it possible to extrapolate the cosmology
consistently for all $0 < w = (R_0/R) < \infty$. For example, in the flat case $\Omega_C = 0$
which our universe seems to approximate\cite{boomerang}, the formula for the expansion rate is

\begin{equation}
\frac{1}{H_0^2} \left( \frac{\dot{R}}{R} \right)^2 = \tilde{\Omega}_M w^3 + \tilde{\Omega}_{\Lambda} w^P
\label{flatexp}
\end{equation}

\noindent This is consistent as a cosmology only if the right-hand side has no zero for a real
positive $w = \hat{w}$. The root $\hat{w}$ is

\begin{equation}
\hat{w} = \left( \frac{ 3(1 - \Omega_M)}{P - 3 \Omega_M} \right)^{\frac{1}{3 - P}}
\label{wcrit}
\end{equation}

\bigskip

\noindent If $0 < \Omega_M < 1$, consistency requires that $P < 3 \Omega_M$. 

\bigskip
\bigskip
\bigskip

\noindent In the more
general case of $\Omega_C \neq 0$ the allowed regions of the $\Omega_M - \Omega_{\Lambda}$
plot for $P = 0,1,2$ are displayed in Fig. 2.

\bigskip
\bigskip

\noindent We see from Eq.(\ref{wcrit}) that if we do violate $P < 3 \Omega_M$ for the flat case
then there is a $\hat{w} > 0$ where the cosmology undergoes a bounce, with $\dot{R} = 0$
and $\dot{R}$ changing sign. This necessarily arises
because of the imposition of $D^{\mu} {\cal T}_{\mu\nu} = 0$ for energy conservation. For this example
it occurs in the past for $\hat{w} > 1$.
The consistency of big bang cosmology back to the time of nucleosynthesis implies
that our universe has not bounced for any $1 < \hat{w} < 10^9$.
It is also possible to construct cosmologies where the bounce occurs in the future!
Rewriting Eq.(\ref{wcrit}) in terms of $\Omega_{\Lambda}$:

\begin{equation}
\hat{w} = \left( \frac{3 \Omega_{\Lambda}}{3 \Omega_{\Lambda} - (3 - P)} \right)^{\frac{1}{3 - P}}
\end{equation}

\bigskip

\noindent If $P < 3$, then any $\Omega_{\Lambda} < 0$ will lead to a solution with $0 < \hat{w} < 1$
corresponding to a bounce in the future. If $P > 3$ the condition for a future
bounce is $\Omega_{\Lambda} < - \left( \frac{P - 3}{3} \right)$. What this means is that
for the flat case $\Omega_C = 0$ with quintessence $P > 0$ it is
possible for the future cosmology to be qualitatively similar to a non-quintessence
closed universe where $\dot{R} = 0$ at a finite future time with a subsequent big crunch.

\bigskip
\bigskip

Another constraint on the cosmological model is provided by nucleosynthesis which requires that
the rate of expansion for very large $w$ does not differ too much from that
of the standard model.

\bigskip

The expansion rate for $P = 0$ coincides for large $w$ with that of the standard model so it
is sufficient to study the ratio:

\begin{eqnarray}
(\dot{R}/R)_P^2/(\dot{R}/R)_{P=0}^2
& 
\stackrel{w \rightarrow \infty}{\rightarrow}
& (3 \Omega_{M} - P)/((3 - P) \Omega_{M})\\
& \stackrel{w \rightarrow \infty}{\rightarrow}& (4 \Omega_{R} - P)/((4 - P)\Omega_{R})
\end{eqnarray}

\noindent where the first limit is for matter-domination and the second is for radiation-domination
(the subscript R refers to radiation).

\bigskip

\noindent The overall change in the expansion rate at the BBN era is therefore

\begin{eqnarray}
(\dot{R}/R)_P^2/(\dot{R}/R)_{P=0}^2
\stackrel{w \rightarrow \infty}{\rightarrow}
(3 \Omega_{M} - P)/((3 - P) \Omega_{M}) \times (4 \Omega^{trans}_{R} - P)/((4 - P)\Omega^{trans}_{R})
\end{eqnarray}
\noindent where the superscript "trans" refers to the transition from radiation domination
to matter domination.
Putting in the values $\Omega_M = 0.3$ and $\Omega^{trans}_R = 0.5$ leads to $P < 0.2$
in order that the acceleration rate at BBN be within 15\% of its value in the standard model,
equivalent to the contribution to the expansion rate at BBN of one chiral neutrino flavor.

\bigskip
\bigskip

\noindent Thus the constraints of avoiding a bounce ($\dot{R} = 0$) in the past, and then requiring
consistency with BBN leads to $0 < P < 0.2$.

\bigskip

We may now ask how this restricted range of $P$ can effect the extraction
of cosmic parameters from observations. This demands an accuracy which has
fortunately begun to be attained with the Boomerang data\cite{boomerang}. If we choose
$l_1 = 197$ and vary $P$ as $P = 0, 0.05, 0.10, 0.15, 0.20$ we
find in the enlarged view of Fig 3 that the variation in the parameters
$\Omega_M$ and $\Omega_{\Lambda}$ can be as large as $\pm 3\%$. To guide
the eye we have added the line for deceleration parameter $q_0 = -0.5$ as suggested
by \cite{Perlmutter,Kirshner}. In the next
decade, inspired by the success of Boomerang (the first paper of true precision cosmology)
surely the sum $(\Omega_M + \Omega_{\Lambda})$ will be examined at much better than $\pm 1\%$
accuracy, and so variation of the exponent of $P$ will provide a useful parametrization
of the quintessence alternative to the standard cosmological model with constant $\Lambda$.

\bigskip

Clearly, from the point of view of inflationary cosmology, the precise
vanishing of $\Omega_C = 0$ is a crucial test and its confirmation
will be facilitated by comparison models such as the present one.

\bigskip

We have also studied the use of the public code CMBFAST\cite{CMBFAST}
and how its normalization compares to that in \cite{FNR}.
For example, with P = 0 and $\Omega_{\Lambda} = 0.3, h_{100} = 0.65$ we
find using CMBFAST that

$\Omega_{\Lambda} = 0.5, l_1 = 284$  ($l_1 = 233$ from \cite{FNR})

$\Omega_{\Lambda} = 0.6, l_1 = 254$  ($l_1 = 208$ from \cite{FNR})

$\Omega_{\Lambda} = 0.7, l_1 = 222$  ($l_1 = 182$ from \cite{FNR})

$\Omega_{\Lambda} = 0.8, l_1 = 191$  ($l_1 = 155$ from \cite{FNR})

The CMBFAST $l_1$ values are consistently $\sim 1.22$ times the $l_1$ values from \cite{FNR}.
As mentioned earlier, this normalization is intermediate between that for
the acoustic horizon ($\sqrt{3}$) and the
photon horizon ($1$).

\bigskip

Finally, we remark that the quintessence model considered here is in
the right direction to ameliorate the "age
problem" of the universe. Taking the age as 14.5Gy for $\Omega_M = 0.3$,
$\Omega_C = 0$ and $h_{100} = 0.65$ the age increases monotonically with $P$.
It reaches slightly over	 15 Gy at the highest-allowed value
$P= 0.2$. This behavior is illustrated in Fig 4
which assumes $\Omega_M = 0.3$ and flatness as $P$ is varied.

\newpage

{\bf Acknowledgments} \hspace{0.5cm}

We thank L.H. Ford and S. Glashow for useful discussions and S. Weinberg for provocative questions.
This work was supported in part by the US Department of Energy
under Grant No. DE-FG02-97ER-41036.

\newpage

\bigskip
\bigskip
\bigskip
\bigskip

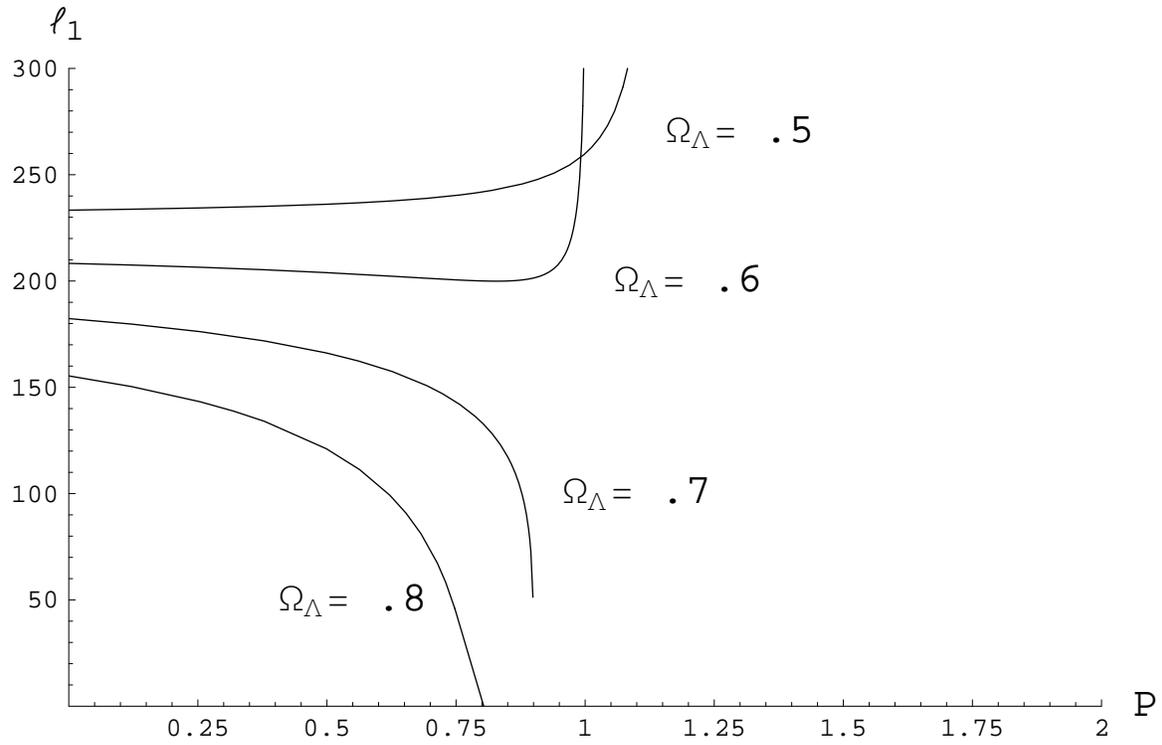
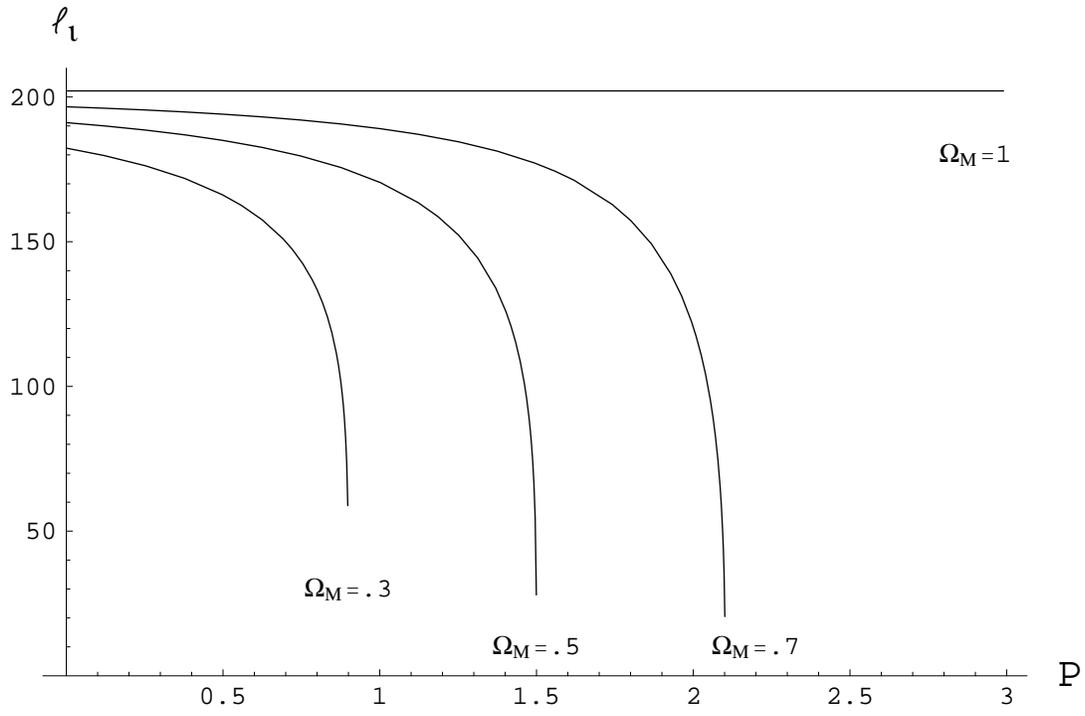
{\bf Figure 1.}

\noindent Dependence of $l_1$ on $P$ for (a) fixed $\Omega_M = 0.3$; (b) fixed $\Omega_C = 0$.

\bigskip

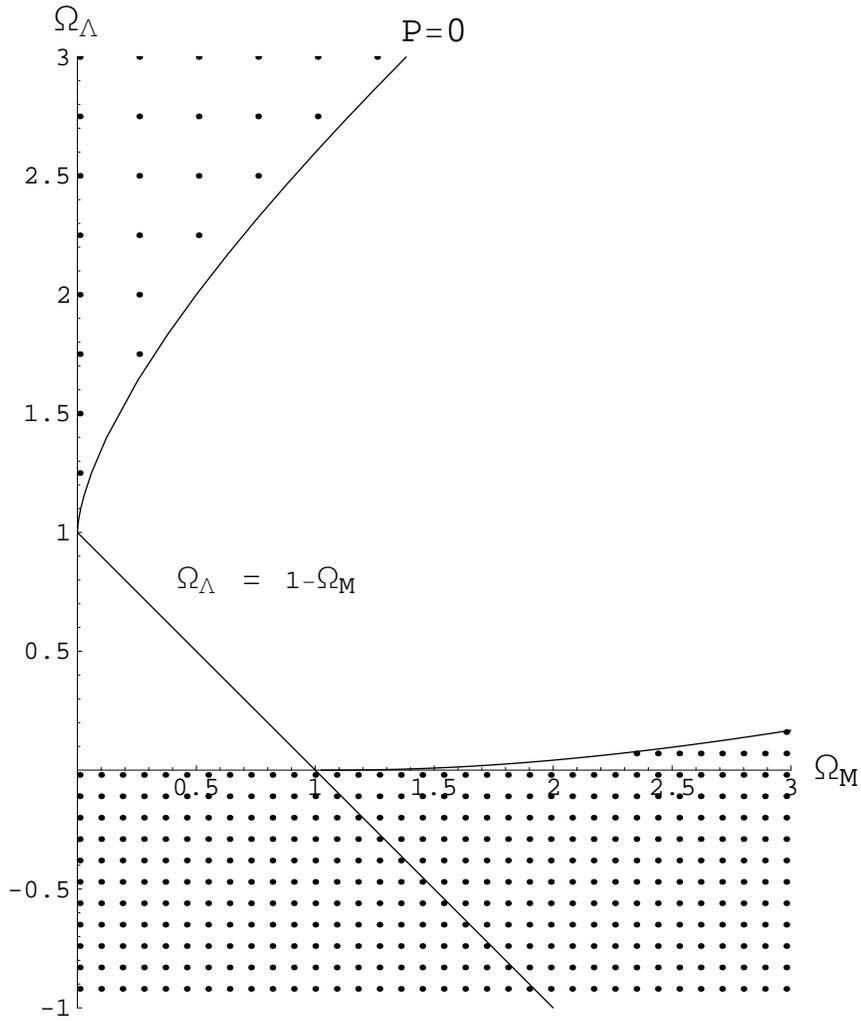
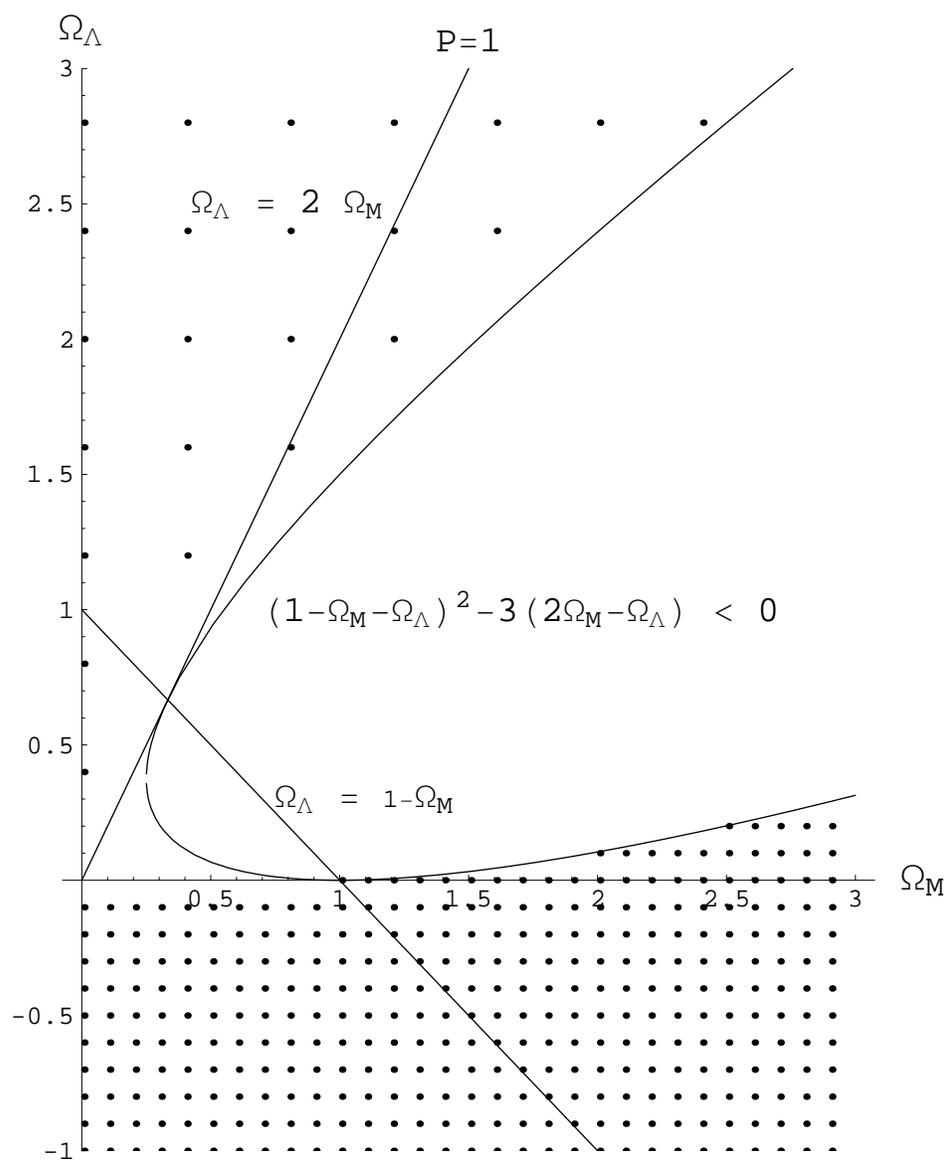
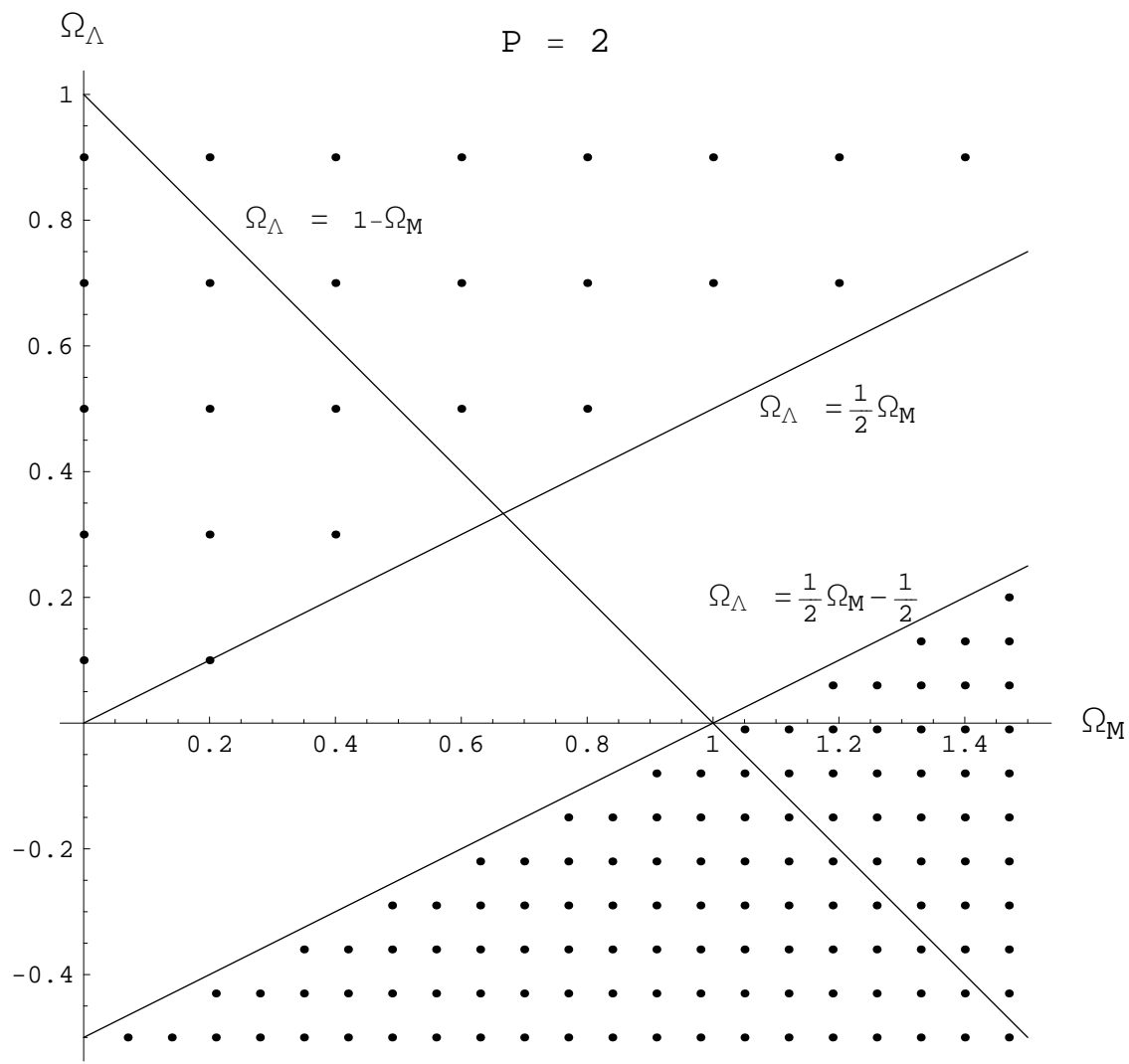
{\bf Figure 2.}

\noindent Regions of the $\Omega_M-\Omega_{\Lambda}$ plot where there
is a future bounce (small dot lattice), no bounce (unshaded) and
a past bounce (large dot lattice) for (a) $P = 0$;
(b) $P = 1$; and (c) $P = 2$.

\bigskip

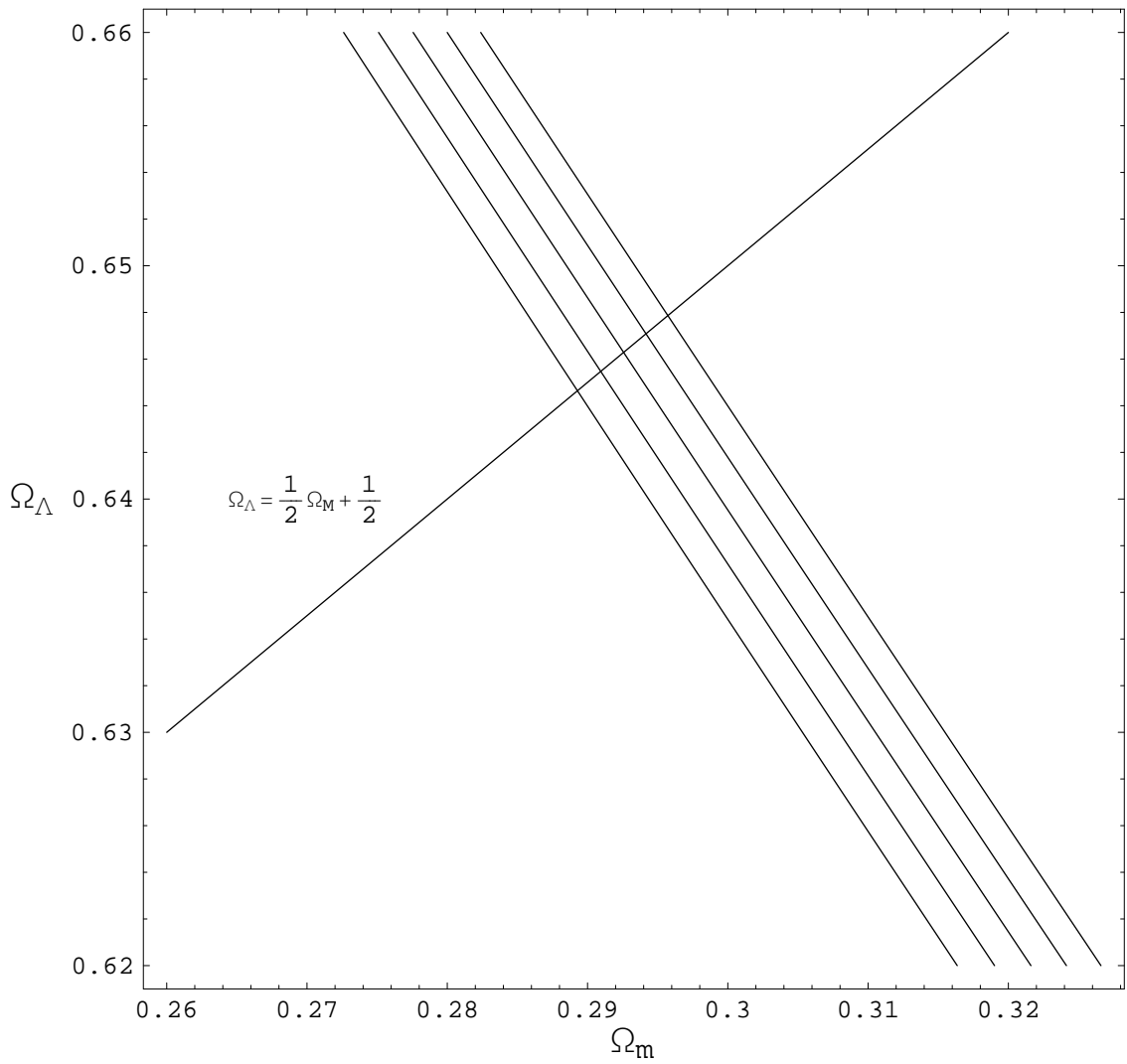
{\bf Figure 3.}

\noindent Enlarged view of $\Omega_M - \Omega_{\Lambda}$ plot to exhibit sensitivity
to $0 \le P \le 0.2$.

\noindent Contours are (right to left) 
$P = 0, 0.05, 0.10, 0.15, 0.20$.

\bigskip

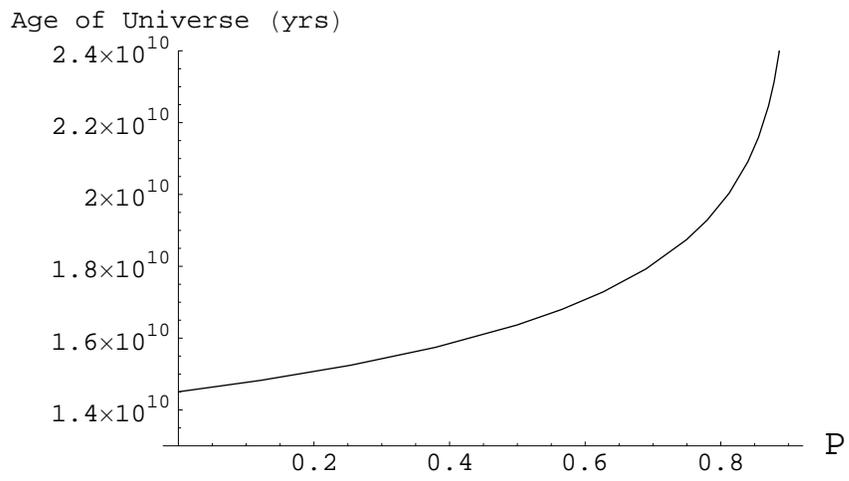
{\bf Figure 4.}

\noindent Age of the universe in units $10^{10} y$ versus $P$.

\noindent This figure assumes $\Omega_M = 0.3$ and flatness $\Omega_C = 0$.

\end{document}